%
%
%
\def\today{\ifcase\month\or January\or February\or March\or April\or May\or
June\or July\or August\or September\or October\or November\or December\fi
\space\number\day, \number\year}
%
%
\newcount\notenumber

\def\note{\global\advance\notenumber by 1 \footnote{$^{\the\notenumber}$}}
%
%
\newif\ifsectionnumbering
\newcount\eqnumber
\def\cleareqnumber{\eqnumber=0}
\def\numbereq{\global\advance\eqnumber by 1
\ifsectionnumbering\eqno(\the\secnumber.\the\eqnumber)\else\eqno
(\the\eqnumber)\fi}
\def\eqalinno{{\global\advance\eqnumber by 1}
\ifsectionnumbering(\the\secnumber.\the\eqnumber)\else(\the\eqnumber)\fi}
\def\name#1{\ifsectionnumbering\xdef#1{\the\secnumber.\the\eqnumber}
\else\xdef#1{\the\eqnumber}\fi}
\def\nosectionnumbering{\sectionnumberingfalse}
\sectionnumberingtrue
%
%
\newcount\refnumber

\immediate\openout1=refs.tex
\immediate\write1{\noexpand\frenchspacing}
\immediate\write1{\parskip=0pt}
\def\ref#1#2{\global\advance\refnumber by 1%
[\the\refnumber]\xdef#1{\the\refnumber}%
\immediate\write1{\noexpand\item{[#1]}#2}}
\def\tie{\noexpand~}

%
%
\font\twelvebf=cmbx10 scaled \magstep1
\newcount\secnumber

\def\newsection#1.{\ifsectionnumbering\cleareqnumber\else\fi%
	\global\advance\secnumber by 1%
	\bigbreak\bigskip\par%
	\line{\twelvebf \the\secnumber. #1.\hfil}\nobreak\medskip\par}
%
%
%
\def \sqr#1#2{{\vcenter{\vbox{\hrule height.#2pt
	\hbox{\vrule width.#2pt height#1pt \kern#1pt
		\vrule width.#2pt}
		\hrule height.#2pt}}}}

%
%
%
\newdimen\fullhsize
\def\fiddle{\fullhsize=6.5truein \hsize=3.2truein}
\def\fullline{\hbox to\fullhsize}
\def\mkhdline{\vbox to 0pt{\vskip-22.5pt
	\fullline{\vbox to8.5pt{}\the\headline}\vss}\nointerlineskip}
\def\mkftline{\baselineskip=24pt\fullline{\the\footline}}
\let\lr=L \newbox\leftcolumn
\def\twocolumns{\fiddle
	\output={\if L\lr \global\setbox\leftcolumn=\columnbox
		\global\let\lr=R \else \doubleformat \global\let\lr=L\fi
		\ifnum\outputpenalty>-20000 \else\dosupereject\fi}}
\def\doubleformat{\shipout\vbox{\mkhdline
		\fullline{\box\leftcolumn\hfil\columnbox}
		\mkftline} \advancepageno}
\def\columnbox{\leftline{\pagebody}}
\nosectionnumbering
\magnification=1200
\def\pr#1 {Phys. Rev. {\bf D#1\tie }}
\def\pe#1 {Phys. Rev. {\bf #1\tie}}
\def\pre#1 {Phys. Rep. {\bf #1\tie}}
\def\pl#1 {Phys. Lett. {\bf #1B\tie }}
\def\prl#1 {Phys. Rev. Lett. {\bf #1\tie }}
\def\np#1 {Nucl. Phys. {\bf B#1\tie }}
\def\ap#1 {Ann. Phys. (NY) {\bf #1\tie }}
\def\cmp#1 {Commun. Math. Phys. {\bf #1\tie }}
\def\imp#1 {Int. Jour. Mod. Phys. {\bf A#1\tie }}
\def\mpl#1 {Mod. Phys. Lett. {\bf A#1\tie}}
\def\tie{\noexpand~}

\parskip=15pt plus 4pt minus 3pt
\headline{\ifnum \pageno>1\it\hfil N=2 Supersymmetry and
Dipole	$\ldots$\else \hfil\fi}
\font\title=cmbx10 scaled\magstep1
\font\tit=cmti10 scaled\magstep1
\footline{\ifnum \pageno>1 \hfil \folio \hfil \else
\hfil\fi}
\raggedbottom


\def\half{{\textstyle{1\over2}}}
\overfullrule0pt
\def\up{\uparrow}
\def\down{\downarrow}
\def\updown{\updownarrow}
\def\zip{0}
\def\mneg{\mskip-\medmuskip}


\rightline{\vbox{\hbox{RU97-08-B}\hbox{hep-th/9711173}}}
\vfill
\centerline{\title N=2 SUPERSYMMETRY AND DIPOLE MOMENTS}
\vfill
{\centerline{\title Ioannis Giannakis
and
James T.~Liu \footnote{$^{\dag}$}
{\rm e-mail: giannak@theory.rockefeller.edu,
jtliu@theory.rockefeller.edu}}
}
\medskip
\centerline{{\tit Physics Department, The Rockefeller
University}}
\centerline{\tit 1230 York Avenue, New York, NY
10021-6399}
\vfill
\centerline{\title Abstract}
\bigskip
{\narrower\narrower
We derive sum rules for the magnetic and electric dipole moments of all
particle states of an $N=2$ supermultiplet.  For short representations,
we find agreement with previously determined $N=1$ sum rules, while
there is added freedom for long representations (expressed as certain
scalar expectation values).  With mild assumptions we find the simple
result that the supersymmetry generated spin adds to the magnetic
(electric) dipole moment with strength corresponding to $g=2$
($g_e=0$).  This result is equally valid for $N=1$, this time without
any further assumptions.
\par}
\vfill\vfill\break


\newsection Introduction.

One of the great successes of the Dirac theory was its
correct prediction of the gyromagnetic ratio of the electron.
This was particularly striking since $g=2$ is twice as large as
expected for the classical orbital motion of a charged point
particle with angular momentum ${{\hbar}/ 2}$. However there
is nothing in Dirac's theory that requires a $g$-value of $2$
for a spin $J={1/ 2}$ particle. Lorentz and gauge invariance
do not prohibit the inclusion of a Pauli term into the
Dirac equation. This term would provide an additional contribution
to the magnetic moment of the electron and alter
the value of $g$. A justification for its
absence is that such a term would render the theory non-renormalizable.
Renormalizability is a matter of asymptotic behaviour at
infinite momentum, so it was not surprising that Weinberg
\ref\wein{S. Weinberg, ``{\it Dynamic and Algebraic Symmetries},''
in: Lectures on elementary particles and quantum field theory,
MIT Press, Cambridge, 1970.}
by demanding good asymptotic behaviour for the
photon forward-scattering amplitudes
showed that $g=2$ for arbitrary
spin charged particles that do not participate in the strong
interactions.  More recently, Ferrara, Porrati and Telegdi
\ref\tel{S. Ferrara,
M. Porrati and V. Telegdi, \pr46 (1992) 3529.}
implemented this particular electromagnetic coupling prescription at
the Lagrangian level. This prescription is different than the minimal
coupling prescription according to which all derivatives $\partial_\mu$
are replaced by covariant ones $D_\mu$ and which yields $g=1/J$
for the gyromagnetic ratio of a particle of spin $J$
\ref\hag{L. Singh and C. Hagen, \pr9 (1974) 898; \pr9 (1974) 910;
F. Belinfante, Phys. Rev. {\bf 92} (1953) 997;
K. Case, Phys. Rev. {\bf 94} (1954) 1442;
C. Hagen and W. Hurley, \prl24 (1970) 1381.}.

The addition of supersymmetry leads to further consequences for the magnetic
dipole moments.  In Ref.~%
\ref\ferrem{S. Ferrara and E. Remiddi, Phys. Lett. {\bf 53B} (1974) 347.}
Ferrara and Remiddi showed that $g=2$ to all orders in perturbation theory
for an $N=1$ chiral multiplet (superspin~0).  On the other hand, when spin-1
fields (superspin~1/2) are involved, Bilchak, Gastmans and Van~Proeyen
\ref\bgvan{C. L. Bilchak, R. Gastmans and A. Van Proeyen, Nucl. Phys.
{\bf B273} (1986) 46.}
showed that supersymmetry does not necessarily demand $g=2$, but
nevertheless leads to a relation between the $g$-factors of the spin-1/2
and spin-1 particles of the superspin~1/2 multiplet.
Subsequently, Ferrara and Porrati
\ref\por{S. Ferrara and M. Porrati, Phys. Lett. {\bf B288} (1992) 85.},
utilizing only the supersymmetry algebra, found exact model independent
sum rules relating the gyromagnetic ratios of all particles within a
single massive $N=1$ supermultiplet.
In particular, for a superspin $j$ multiplet (with
particles of spins $j-1/2$, $j$, $j$, and $j+1/2$), the gyromagnetic ratios
may be expressed in terms of a single free parameter, namely the transition
moment between the spin $j-1/2$ and $j+1/2$ states of the multiplet.
Furthermore, when this non-diagonal moment vanishes, the sum rule simply
states that $g=2$ for all members of the supermultiplet.  In particular,
this confirms the result of [\ferrem] that $g=2$ for a chiral multiplet
since there is no room for a transition moment for superspin 0.  Thus
the anomalous magnetic moment of the electron (in a supersymmetric
standard model) identically vanishes, as long as supersymmetry remains
unbroken.  This is but one of the manifestations of how supersymmetry
alone provides powerful results independent of any particular model.

More recently, dipole moment sum rules have been applied in order to test
the conjectured equivalence of string states and black holes.  For this
conjecture to be true, not only do masses, charges and representations have
to agree, but so do other physical properties such as electric and magnetic
dipole moments.  Furthermore, it was anticipated in
\ref\mdjr{M. J. Duff and J. Rahmfeld, \np 481 (1996) 332.}
that examination of dipole moments would shed further light on the bound
state picture of black holes and $p$-branes.  An extensive study of
dipole moments for strings and black holes in an $N=4$ context
found complete agreement between the gyromagnetic ratios of states in
both short and intermediate multiplets
\ref\jim{M. J. Duff, J. T. Liu and J. Rahmfeld, \np494 (1997) 161.}.
However in that work it was realized that the gyromagnetic ratios for short
multiplets are completely determined based on supersymmetry alone.  Thus,
as long as $N=4$ supersymmetry is unbroken, the $g$-factors must necessarily
agree between corresponding supersymmetric black holes and $N_R=1/2$
heterotic string states, and hence do not provide a true test of the strings
as black holes conjecture.  On the other hand, supersymmetry becomes a lot
less restrictive for intermediate and long multiplets.  While the
correspondence between $N_R>1/2$ states and non-extremal black holes is not
so clear, application of $T$-duality allowed a comparison of $g$-factors for
intermediate black holes and corresponding Type II string states, where
agreement was found [\jim].

This issue of how much freedom is actually present in the gyromagnetic
ratios has motivated us to examine both electric and magnetic dipole moment
sum rules in a more general extended supersymmetry context.  Thus in the
following we extend the results of [\por] and derive completely general
$N=2$ dipole sum rules for particles in either short or long multiplets
of $N=2$ supersymmetry.  We find the interesting result that, contrary to
expectations, the $N=2$ sum rules are weaker than the $N=1$ case in that
they depend on additional quantities (certain scalar expectation values)
beyond just the mass, electric charge and central charge of the
representation.  This additional freedom disappears under certain mild
assumptions, in which case the $N=2$ sum rules become a simple
generalization of the $N=1$ case.  In particular, we find that the
supersymmetry generated spin adds to the magnetic (electric) dipole moment
with a factor of $g=2$ ($g_e=0$).

In the next section we set our notations
by discussing the $N=2$ supersymmetry algebra and its
representations, while in section $3$ we briefly discuss
the linear multiplet of $N=2$ supersymmetry. Finally,
in section $4$ we derive model
independent sum rules for the gyromagnetic and gyroelectric
ratios of the members of an $N=2$ supermultiplet.  These sum rules are
presented in both a completely general fashion, and also in simplified form
whenever the assumptions mentioned above are valid.  Concluding remarks are
presented in section 5.


\newsection N=2 Supersymmetry Algebra.

Our starting point is the $N=2$ supersymmetry algebra, which admits a single
complex central charge $Z=U+iV$.  Using the $N=2$ Majorana condition
${\overline Q}^{i}=i{\epsilon^{ij}}Q^{T}_{j}(C{\gamma_{5}})$, the algebra
may be expressed as
$$
\lbrace Q_{\alpha i}, Q_{\beta j} \rbrace=-2i({\gamma^{\mu}}
{\gamma_5}C)_{\alpha\beta}{\epsilon}_{ij}P_{\mu}
+ i{\epsilon}_{ij}({\gamma_5}C)_{\alpha\beta}U
- {\epsilon}_{ij}C_{\alpha\beta}V,
\numbereq\name{\eqvare}
$$
where $i,j=1,2$ are $SU(2)$ indices and
$C$ is the charge conjugation matrix obeying
$C{\gamma^{\mu}}C^{-1}=-{\gamma^{\mu T}}$ and $C^2=-1$.  For a massive
single particle state, we may work in the rest frame $P^{\mu}=(M,0,0,0)$.
Defining chiralities
$$
{\gamma_5}Q^{L}_{i}=-Q^{L}_{i}, \qquad
{\gamma_5}Q^{R}_{i}=Q^{R}_{i},
\numbereq
$$
and helicities
$$
{\gamma^{12}}Q_{\pm{1\over 2}i}={\mp i}Q_{\pm {1\over 2}i},
\numbereq
$$
the supersymmetry algebra can be recast as follows:
$$
\eqalign{
&\lbrace Q^{L}_{\pm{1\over 2}1}, Q^{L}_{\mp{1\over 2}2} \rbrace=
{\mp}i Z, \cr
&\lbrace Q^{R}_{\pm{1\over 2}1}, Q^{L}_{\mp{1\over 2}2} \rbrace=
-2iM,\cr} \qquad
\eqalign{
&\lbrace Q^{R}_{\pm{1\over 2}1}, Q^{R}_{\mp{1\over 2}2} \rbrace=
{\pm}i\overline{Z},\cr
&\lbrace Q^{L}_{\pm{1\over 2}1}, Q^{R}_{\mp{1\over 2}2} \rbrace=
2iM, \cr}
\numbereq\name{\eqana}
$$
indicating the expected splitting between mass and central charge terms
in a Weyl basis%
\note{To fix our phase conventions, we work in the Dirac
representation for the ${\gamma}$-matrices and take
$\gamma_5=i{\gamma^{0}}{\gamma^{1}} {\gamma^{2}}{\gamma^{3}}$ and
$C=i{\gamma^{0}}{\gamma^{2}}$.  The spinors then decompose as
${\sqrt 2}Q_{\alpha i}^T=Q^{L}_{{1\over 2}i}[1\ 0\ 1\ 0]
+Q^{L}_{-{1\over 2}i}[0\ 1\ 0\ 1]
+Q^{R}_{{1\over 2}i}[-1\ 0\ 1\ 0]
+Q^{R}_{-{1\over 2}i}[0\ 1\ 0\ -1]$.}.

The above $N=2$ algebra may be diagonalized by introducing the linear
combinations
$$
Q^{\pm}_{{1\over 2}i}={1\over {\sqrt 2}} \lbrack Q^{L}_{{1\over 2}i}
{\mp}ie^{i{\alpha}}Q^{R}_{{1\over 2}i} \rbrack, \qquad
Q^{\pm}_{-{1\over 2}i}={1\over {\sqrt 2}} \lbrack {\pm}iQ^{R}_{-{1\over 2}i}
-e^{-i{\alpha}}Q^{L}_{-{1\over 2}i} \rbrack,
\numbereq\name{\eqkara}
$$
where $\alpha$ is the phase of the central charge, $Z=e^{i\alpha}|Z|$.
In terms of these mixed chirality supercharges, the algebra (\eqana) now
takes the simple form
$$
\lbrace Q^{+}_{\pm{1\over 2}1}, Q^{+}_{\mp{1\over 2}2} \rbrace= 2M+|Z|, \qquad
\lbrace Q^{-}_{\pm{1\over 2}1}, Q^{-}_{\mp{1\over 2}2} \rbrace= 2M-|Z|,
\numbereq\name{\eqgeor}
$$
indicating explicitly the $N=2$ Bogomol'nyi bound, $2M\ge|Z|$.  Massive
representations thus split up into either long or short multiplets, with the
latter corresponding to saturation of the Bogomol'nyi bound, $2M=|Z|$.

For a long representation, we may rescale the supercharges according to
$q^{\pm}_{{\pm}{1\over 2}i}=
(2M{\pm}|Z|)^{-{1\over 2}}Q^{\pm}_{{\pm}{1\over 2}i}$ to
recover the Clifford algebra for four fermionic degrees of freedom.
{}From the form of this algebra it follows that one can construct
its irreducible representations by starting with a superspin $j$
Clifford vacuum, $|j\rangle$, annihilated by $q^{\pm}_{{\pm}{1\over 2}2}$,
and acting on it with the creation operators $q^{\pm}_{{\pm}{1\over 2}1}$.
As a result, we see that the long representation has dimension
$(2j+1)\times2^4$ where $2j+1$ is the degeneracy of the original spin $j$
state.  The spins of the states are given by the addition of angular
momenta, $j \times [(1) + 4(1/2) + 5(0)]$, giving generically
states of spins $j-1$ to $j+1$ with degeneracies $1,4,5+1,4,1$
(provided $j\ge 1$).

When the Bogomol'nyi bound is saturated, $2M=|Z|$, the supercharges
$Q^{-}_{{\pm} {1\over 2}i}$ are represented trivially
and the algebra becomes the algebra of two fermionic
annihilation and creation operators.  The short representations thus
contain spins $j \times [(1/2) + 2(0)]$ (generically giving spins $j-1/2$
to $j+1/2$ with degeneracies $1,2,1$) and have dimension $(2j+1)\times2^2$.
These short multiplets of $N=2$ supersymmetry correspond to
the same particle content as massive supermultiplets of $N=1$, and in fact
satisfy identical dipole moment sum rules, as will be demonstrated below.

Since the supercharges $Q^{\pm}_{{1\over 2}1},
Q^{\pm}_{-{1\over 2}1}$ are operators of spin $1\over 2$, this leads to a
natural shorthand notation for labeling the states of a generic long $N=2$
multiplet.  We denote the superspin $j$ Clifford vacuum by $|00\rangle$
where the first (second) entry corresponds to the action of the $2M+|Z|$
($2M-|Z|$) normalized creation/annihilation algebra of Eqn.~(\eqgeor).  Acting
on this state with the normalized supercharges $q^{+}_{{1\over 2}1}$ or
$q^{+}_{-{1\over 2}1}$ then results in the spin `up' or `down' states
$|{\uparrow} 0\rangle$ or $|{\downarrow} 0\rangle$ respectively.
On the other hand, the action of $q^{-}_{{1\over 2}1}$ or
$q^{-}_{-{1\over 2}1}$ results in the states $|0 {\uparrow}\rangle$ or
$|0 {\downarrow}\rangle$.  The action of several $q$'s on the Clifford
vacuum are then represented in a similar manner.  For example the action
of all four supercharges is denoted by $|{\updownarrow}{\updownarrow}\rangle$.
Note that states in a short multiplet will always have a 0 in the second
entry.  Finally, it should be kept in mind that the physical states of
the supermultiplet correspond to the addition of angular momentum $j$ to the
above spin states using the appropriate Clebsch-Gordan coefficients.

\newsection Conserved Currents in Supersymmetric Theories.

For $N=1$ supersymmetry, any conserved current commuting
with the supersymmetry generators must belong to a real linear
multiplet.  In the present case this generalizes to a $N=2$ linear multiplet
\ref{\soh}{M. Sohnius, \pl81 (1979) 8;
D. Breitenlohner and M. Sohnius, \np165 (1980) 483.}
consisting of $(K^{\alpha}, \xi_{i}, S, P, J_{\mu})$ where $K^\alpha$ is a
$SU(2)$ triplet scalar, $\xi_i$ is a $SU(2)$ doublet Majorana spinor and
$S, P$ are real scalars.  We take the linear multiplet to transform without
central charge, so that the current is conserved, $\partial^{\mu}J_{\mu} =0$.
As a result the multiplet includes $8$ bosonic and $8$
fermionic degrees of freedom. The transformation properties of the
components under a supersymmetry variation are given by
$$
\eqalign{
{\delta}K^{\alpha}&=-{\overline{\epsilon}}^{i}{\sigma^{\alpha}_{ij}}
{\xi_j},\cr
{\delta}S&=-{\overline{\epsilon}}^{i}{\gamma^{\mu}}{\partial_{\mu}}
{\xi_i},\cr}
\qquad
\eqalign{
{\delta}{\overline{\xi_i}}&=-{\overline{\epsilon}}^{i}
(S+{\gamma_5}P-{\gamma^{\mu}}J_{\mu})+{\overline{\epsilon}}^{j}
{\sigma^{\alpha}_{ji}}{\gamma^{\mu}}{\partial_{\mu}}K^{\alpha},\cr
{\delta}P&={\overline{\epsilon}}^{i}{\gamma^{\mu}}
{\gamma_5}{\partial_{\mu}}{\xi_i},
\qquad
{\delta}J_{\mu}=-{\overline{\epsilon}}^{i}{\gamma_{\mu\nu}}{\partial^{\nu}}
{\xi_i}.\cr}
\numbereq\name{\eqgian}
$$
It follows that two successive supersymmetry transformations on the
conserved current $J_{\mu}$ give
$$
{\delta}_{\eta}{\delta}_{\epsilon}J_{\mu}=
i{\overline{\epsilon}}^{i}{\gamma_{\mu\nu}}{\partial^{\nu}}\lbrack
(S+{\gamma_5}P-{\gamma^{\lambda}}J_{\lambda}){\delta_{i}^{j}}
+i{\gamma^{\lambda}}{\partial_{\lambda}}K^{\alpha}{\sigma^{\alpha}
_{ij}} \rbrack{\eta}_{j}.
\numbereq\name{\eqpasa}
$$
The matrix elements of this equation between states which belong to the
same $N=2$ multiplet give rise to sum rules for the gyromagnetic
and gyroelectric ratios of the particle states.

In order to derive both electric and magnetic dipole sum rules, we need the
following expansions for the matrix elements of $J_{\mu}$:
$$
\eqalign{
\langle j', m', \vec p\,|J_{0}|j, m, 0\rangle&=
2Me_{j}{\delta_{jj'}}{\delta_{mm'}}
+2Mp_{i}\langle j', m', 0|d^{i}|j, m, 0\rangle+O(p^2),\cr
\langle j', m', \vec p\,|J_{i}|j, m, 0\rangle&=
-e_{j}p_{i}{\delta_{jj'}}{\delta_{mm'}}
-2iM{\epsilon_{ijk}}p_{j}\langle j', m', 0|{\mu}^{k}|j, m, 0\rangle
+O(p^2).\cr}
\numbereq\name{\eqnini}
$$
When $J_{\mu}$ is the electromagnetic current, $e, \vec d, \vec \mu$
are the electric charge, the electric dipole moment and the
magnetic dipole moment respectively.  Our notation follows [\por], where
$|j,m,\vec p\,\rangle$ corresponds to a single particle state of spin $j$,
$z$-component of spin $m$ and 3-momentum $\vec p$.  We emphasize that the
expansion of the matrix elements of the current in powers of
the momenta is based solely on current conservation.  As a convenience,
whenever $j$ and $m$ are not explicitly needed, we use the shorthand
notation $|j,m,\vec p\,\rangle = |\alpha,\vec p\,\rangle$.

\newsection Derivation of the $N=2$ sum rules.

In [\por], the $N=1$ magnetic moment sum rule was derived by noting that a
generic double supersymmetry variation may be expressed as
$$
\eqalign{
\delta_\eta\delta_\epsilon \hat{\cal O} &=
[\overline{\eta} Q, [\overline{\epsilon}Q, \hat{\cal O}]]\cr
&=\overline{\eta}Q\overline{\epsilon}Q\hat{\cal O}
-\overline{\eta}Q\hat{\cal O}\overline{\epsilon}Q
-\overline{\epsilon}Q\hat{\cal O}\overline{\eta}Q
+\hat{\cal O}\overline{\epsilon}Q\overline{\eta}Q.\cr}
\numbereq\name{\eqsamba}
$$
Evaluating this expression between given single particle states
$\langle\alpha|$ and $|\beta\rangle$, and noting that the supercharge
$Q$ generates superpartners ($Q|\alpha\rangle \sim |\tilde\alpha\rangle$),
we are then able to relate matrix elements of $\hat{\cal O}$ between
different states of a supermultiplet, provided $\delta_\eta\delta_\epsilon
\hat{\cal O}$ is known.  The magnetic dipole sum rules then follow by
choosing $\hat{\cal O}$ to be the conserved current $J_\mu$, and using
(\eqnini) to determine its matrix elements.

This general procedure is simplified in practice by choosing the global
supersymmetry transformation parameters $\eta$ and $\epsilon$ in such a
way so that several
terms on the right hand side of (\eqsamba) act as annihilation operators on
the initial or final states and hence may be dropped.  In particular, to
lowest order in momentum $\vec p$, and making use of the Lorentz boost
operator $|\alpha, \vec p\,\rangle = L(\vec p)|\alpha, 0\rangle$, we find
$$
\eqalign{
\langle\alpha,\vec p\,|\delta_\eta\delta_\epsilon J_\mu|\beta, 0\rangle=&\>
\langle\alpha,\vec p\,|J_\mu\overline{\epsilon}Q\overline{\eta}Q|\beta,0
\rangle
-\langle\alpha,0|\overline{\epsilon}QL^{-1}(\vec p)J_\mu\overline{\eta}Q
|\beta,0\rangle\cr
&-{\delta}_{\mu0}{{p^i}\over 2M}\langle\alpha,0|
\overline{\epsilon}\gamma^{0i}QJ_{0}\overline{\eta}Q|\beta,0\rangle
+O(p^2),\cr}
\numbereq\name{\eqasxuti}
$$
provided $\langle\alpha,\vec p\,|\overline{\eta}Q=0$.  Note that we have used
the fact that $Q$ transforms as a spinor so that
$[L^{-1}(\vec p), \overline{\eta}Q]={1\over 2}p^i
\overline{\eta}\gamma^{0i}Q+O(p^{2})$.  This expansion of the Lorentz boost
gives rise to the last term above, which only contributes to $J_0$ matrix
elements (and hence is only important in deriving the electric dipole moment
sum rules).

Turning to the left hand side of (\eqasxuti), and using the double
supersymmetry variation (\eqpasa), we find
$$
\eqalign{
{\delta}_{\eta}{\delta}_{\epsilon}J_i&=
-i{\epsilon_{ijk}}p^{j}
{\overline{\epsilon}}\lbrack
{\gamma_{\kappa}}{\gamma^5}J_{0}+
{\gamma_{\kappa}}{\gamma^{0}}{\gamma^5}S+i{\gamma_{\kappa}}{\gamma^0}P
\rbrack{\eta}+O(p^2),\cr
\delta_\eta\delta_\epsilon J_0 &=
-p_i\overline{\epsilon}[\gamma^i J_0 -
\gamma^i\gamma_0S-i\gamma^i\gamma_0\gamma^5P]\eta+O(p^2).\cr}
\numbereq\name{\eqpadxz}
$$
Note that, while the auxiliary field $K^\alpha$ is unimportant at this
order, matrix elements of $S$ and $P$ remain and cannot be ignored.
Since the conserved current multiplet commutes with supersymmetry, these
matrix elements, like the electric charge, are identical for all states in a
given representation.  Thus we may define $S$ and $P$ expectations as
$$
\eqalign{
\langle\alpha, 0|S|\beta,0\rangle &= 2M{\cal S}\delta_{\alpha\beta},\cr
\langle\alpha, 0|P|\beta,0\rangle &= 2M{\cal P}\delta_{\alpha\beta}.\cr
}
\numbereq\name{\eqvcxiu}
$$

Since a generic long multiplet of $N=2$ contains many more states than that
of $N=1$, we find it convenient to take a systematic approach to examining
the matrix elements of (\eqasxuti) on various states.  In particular, the
magnetic dipole moment sum rules may be derived in two parts: $i)$ a set of
``vanishing sum rules'' concerning elements of the dipole moment operator
between different states of the multiplet, and $ii)$ ``diagonal sum rules''
relating diagonal elements of different states.

Derivation of the vanishing sum rules follows by choosing the parameters of
transformation to satisfy $\langle{\alpha}, \vec p\,|{\overline{\eta}}Q
=\langle{\alpha}, \vec p\,|{\overline{\epsilon}}Q=0$, in which case
Eqn.~(\eqasxuti) becomes
$$
\langle{\alpha}, \vec p\,|J_i{\overline{\epsilon}}Q{\overline{\eta}}Q
|{\beta}, 0\rangle
=
\langle\alpha, \vec p\,|\delta_\eta\delta_\epsilon J_i|\beta, 0\rangle
+O(p^2).
\numbereq\name{\eqtsour}
$$
By choosing both $\langle{\alpha}, \vec p\,|$ and $|\beta,0\rangle$ to denote
different states of the supermultiplet, this allows us to compute the
off-diagonal matrix elements of $\vec\mu$ in terms of the charges $e$,
${\cal S}$ and ${\cal P}$ that show up in the double variation on the
right hand side.

The diagonal sum rules are derived instead by taking states satisfying
$\langle{\alpha}, \vec p\,|{\overline{\eta}}Q =
{\overline{\epsilon}}Q|{\beta}, 0\rangle =0$.  For this case, we find
$$
\langle\alpha,0|\overline{\epsilon}QL^{-1}(\vec p)J_i\overline{\eta}Q
|\beta,0\rangle = \langle\alpha,\vec p\,|J_i [\overline{\epsilon}Q,
\overline{\eta}Q]|\beta,0\rangle
-\langle\alpha,\vec p\,|\delta_\eta\delta_\epsilon J_i|\beta,0\rangle
+O(p^2).
\numbereq\name{\eqwasind}
$$
Note that $[\overline{\epsilon}Q,\overline{\eta}Q]$ corresponds to
the supersymmetry algebra, and hence gives $2M\pm|Z|$ for appropriate
parameters.  Thus, when generating properly normalized superpartners,
the above expression simply states that the dipole moment of the
superpartner (on the left) is the same as the dipole moment of the original
state (on the right) with the addition of a supersymmetry generated
correction given by $\delta_\eta\delta_\epsilon J_i$.

We recall that a basic long multiplet contains 16 states, divided into 8+8
based on integer or half-integer spins.  Since the magnetic dipole operator
(being a vector) does not connect integer and half-integer spins, its matrix
elements on these 16 states split up into two $8\times8$ block diagonal
pieces.  Applying both vanishing and diagonal sum rules, the matrix elements
of the $z$-component of $\vec\mu$, $\langle\alpha,0|\mu^3|\beta,0\rangle$,
are given in Tables 1 (integer spins) and 2 (half-integer spins).
\vadjust{\topinsert
$$
\kern-11pt
\matrix{
\left\langle\zip\zip\right|\cr
\left\langle\up\up\right|\cr\left\langle\down\down\right|\cr
\left\langle\up\down\right|\cr\left\langle\down\up\right|\cr
\left\langle\updown\mneg\zip\right|\cr\left\langle\zip\mneg\updown\right|\cr
\left\langle\updown\updown\right|\cr}
\left[\matrix{
\mu_0\cr
&\mu_0+\alpha^++\alpha^-\cr
&&\mu_0-\alpha^+-\alpha^-\cr
&&&\mu_0+\alpha^+-\alpha^-&&-i\tilde w&-i\tilde w\cr
&&&&\mu_0-\alpha^++\alpha^-&-i\tilde w&-i\tilde w\cr
&&&i\tilde w&i\tilde w&\mu_0\cr
&&&i\tilde w&i\tilde w&&\mu_0\cr
&&&&&&&\mu_0\cr}\right]
\kern-11pt
$$
\smallskip
\centerline{Table 1: Matrix elements of $\mu^3$ on the integer spin states
of a long multiplet.}
\endinsert}%
\vadjust{\topinsert
$$
\kern-3pt
\matrix{
\left\langle\up\mneg\zip\right|\cr\left\langle\zip\mneg\up\right|\cr
\left\langle\down\mneg\zip\right|\cr\left\langle\zip\mneg\down\right|\cr
\left\langle\up\updown\right|\cr\left\langle\updown\up\right|\cr
\left\langle\down\updown\right|\cr\left\langle\updown\down\right|\cr}
\left[\matrix{
\mu_0+\alpha^+&-i\tilde w\cr
i\tilde w&\mu_0+\alpha^-\cr
&&\mu_0-\alpha^+&i\tilde w\cr
&&-i\tilde w&\mu_0-\alpha^-\cr
&&&&\mu_0+\alpha^+&i\tilde w\cr
&&&&-i\tilde w&\mu_0+\alpha^-\cr
&&&&&&\mu_0-\alpha^+&-i\tilde w\cr
&&&&&&i\tilde w&\mu_0-\alpha^-\cr}\right]
\kern-3pt
$$
\smallskip
\centerline{Table 2: Matrix elements of $\mu^3$ on the half-integer spin
states of a long multiplet.}
\endinsert}%
We have taken, by definition, $\mu_0=\langle\zip\zip|\mu^3|\zip\zip\rangle$.
The real numbers $\alpha^\pm$ and $\tilde w$ are given by
${\alpha^{+}}={(e+v)\over {2M+|Z|}}$,
${\alpha^{-}}={(e-v)\over {2M-|Z|}}$ and
$\tilde w = {w\over\sqrt{4M^2-|Z|^2}}$, where $v$ and $w$ are the rotated
scalar expectation values
$$
\pmatrix{v\cr w} =
\pmatrix{\cos\alpha&\sin\alpha\cr -\sin\alpha&\cos\alpha}
\pmatrix{{\cal S}\cr {\cal P}}.
\numbereq
$$
This is the main result of our paper. We note that the
matrix elements of the dipole moment operator between
the different states of a long $N=2$ supermultiplet
are expressed in terms of the mass $M$, the
central charge $|Z|$ and the charges $e$, ${\cal S}$, ${\cal P}$.

Short multiplets, on the other hand, are expected to behave as massive
$N=1$ multiplets. The relation $2M=|Z|$ which holds
for short multiplets implies that the supercharges
$Q^{-}_{\pm{1\over 2}i}$ are represented trivially
and as a result all the states with up or down arrows
in the second entry disappear.  Then, by picking $\epsilon$ and
$\eta$ in Eqn.~(\eqwasind) to select the $Q^-_{\pm{1\over2} i}$
supercharges, we are left with $\langle\alpha,\vec p\,|
\delta_\eta\delta_\epsilon J_i|\beta,0\rangle=0+O(p^2)$.  Combined with the
explicit supersymmetry variation, (\eqpadxz), this gives rise to relations
between $e$, ${\cal S}$, ${\cal P}$, $M$ and $Z$. More specifically we
find that $e=v$ and $w=0$.  These two relations can then be written as one:
$$
{\cal S}+i{\cal P}={eZ\over 2M}.
\numbereq\name{\eqazoicp}
$$
The matrix elements of $\mu^3$ on the states of a short
multiplet simplify as follows
$$
\matrix{
\left\langle\zip\zip\right|\cr
\left\langle\up\mneg\zip\right|\cr\left\langle\down\mneg\zip\right|\cr
\left\langle\updown\mneg\zip\right|\cr}
\left[\matrix{
\mu_0\cr
&\mu_0+{e\over 2M}\cr
&&\mu_0-{e\over 2M}\cr
&&&\mu_0\cr}\right],
\numbereq
$$
in agreement with the results of [\por].

In a similar manner we can derive sum rules for the
electric dipole moments. The results are summarized in Tables 3 and 4,
\vadjust{\topinsert
$$
\kern-17pt
\matrix{
\left\langle\zip\zip\right|\cr
\left\langle\up\up\right|\cr\left\langle\down\down\right|\cr
\left\langle\up\down\right|\cr\left\langle\down\up\right|\cr
\left\langle\updown\mneg\zip\right|\cr
\left\langle\zip\mneg\updown\right|\cr
\left\langle\updown\updown\right|\cr}
\left[\matrix{
d_0\cr
&d_0+iw^+-iw^-\cr
&&d_0-iw^++iw^-\cr
&&&d_0+iw^++iw^-&&\tilde u&\tilde u\cr
&&&&d_0-iw^+-iw^-&\tilde u&\tilde u\cr
&&&-\tilde u&-\tilde u&d_0\cr
&&&-\tilde u&-\tilde u&&d_0\cr
&&&&&&&d_0\cr}\right]
\kern-17pt
$$
\smallskip
\centerline{Table 3: Matrix elements of $d^3$ on the integer spin states
of a long multiplet.}
\endinsert}%
\vadjust{\topinsert
$$
\kern-17pt
\matrix{
\left\langle\up\mneg\zip\right|\cr\left\langle\zip\mneg\up\right|\cr
\left\langle\down\mneg\zip\right|\cr\left\langle\zip\mneg\down\right|\cr
\left\langle\up\updown\right|\cr\left\langle\updown\up\right|\cr
\left\langle\down\updown\right|\cr\left\langle\updown\down\right|\cr}
\left[\matrix{
d_0+iw^+&\tilde u\cr
-\tilde u&d_0-iw^-\cr
&&d_0-iw^+&-\tilde u\cr
&&\tilde u&d_0+iw^-\cr
&&&&d_0+iw^+&-\tilde u\cr
&&&&\tilde u&d_0-iw^-\cr
&&&&&&d_0-iw^+&\tilde u\cr
&&&&&&-\tilde u&d_0+iw^-\cr}\right]
\kern-17pt
$$
\smallskip
\centerline{Table 4: Matrix elements of $d^3$ on the half-integer spin
states of a long multiplet.}
\endinsert}%
where $d_0 = \langle00|d^3|00\rangle$ and $w^\pm={w\over2M\pm|Z|}$ and
$\tilde u={(e|Z|/2M)-v\over\sqrt{4M^2-|Z|^2}}$.  Curiously enough, we see that
in general the electric dipole moments are non-vanishing, even with $d_0=0$.
Demanding that $N=2$ supersymmetry does not generate an electric dipole
moment when none was initially present then requires $0=w^+=w^-=\tilde u$,
so that in fact $v={e|Z|\over2M}$ and $w=0$, corresponding to the condition
(\eqazoicp) that was found for short multiplets.

While in general we have been unable to ascertain whether or not
Eqn.~(\eqazoicp) must continue to hold for long multiplets, it appears that
this condition is true in practice for many explicit $N=2$ realizations.
In fact, both magnetic and electric dipole moment sum rules greatly
simplify whenever Eqn.~(\eqazoicp) is valid.  To see this, note that in this
case $\alpha^+=\alpha^-={e\over2M}$, so that the magnetic dipole
matrix elements of Tables 1 and 2 become diagonal, with the addition of
${e\over2M}$ units of dipole moment for every $\half$ unit of spin generated
by the supersymmetry algebra.  The electric dipole matrix elements of Tables
3 and 4 are even simpler; they only contain the original electric dipole
moment $d_0$, with no addition from supersymmetry.

Finally, we present the $g$-factor sum rules for the physical states of the
superspin $j$ multiplet by adding the supersymmetry generated spin to the
original spin $j$ using appropriate Clebsch-Gordon combinations%
\note{While the addition of angular momentum was an integral part of the
derivation of the $N=1$ sum rule [\por], we find it more convenient to keep
the superpartner generation and the Clebsch-Gordon manipulation separate,
especially for large multiplets.}.
Recalling that the states of the $N=2$ long multiplet are generated by
$j\times [(1)+4(1/2)+5(0)]$, we need the Clebsch-Gordon coefficients for
$j\times 1$ and $j\times 1/2$.  For example, for the latter, we use
$$
\eqalign{
|j+\half , m+\half\rangle&
={1\over {{\sqrt{2j+1}}}}
\big[ {\sqrt{j+m+1}}|j,m;\half, \half\rangle
+{\sqrt{j-m}}|j,m+1;\half, -\half\rangle \big], \cr
|j-\half, m+\half\rangle&={1\over {{\sqrt{2j+1}}}}
\big[ -{\sqrt{j-m}}|j,m;\half, \half\rangle
+{\sqrt{j+m+1}}|j,m+1;\half, -\half\rangle \big]. \cr}
\numbereq\name{\eqivic}
$$
The $g$-factors may then be defined in terms of the matrix elements of
the magnetic dipole moment operator $\vec\mu$ between states of definite
angular momentum using the Wigner-Eckart theorem as follows:
$$
\langle j, m|{\mu_3}|j, m\rangle={e\over {2M}}mg_j,
\numbereq\name{\eqofori}
$$
where $j$ and $m$ label any state of angular momentum $j$ and $z$-component
$m$.  Combining Eqns.~(\eqivic) and (\eqofori), and using the matrix
elements of Table~2, then gives for the four $j+1/2$ and four $j-1/2$
states
$$
g_{j+{1\over 2}}=g_{j}+{{g_{s}-g_{j}}\over {2j+1}},\qquad
g_{j-{1\over 2}}=g_{j}-{{g_{s}-g_{j}}\over {2j+1}},
\numbereq\name{\eqdsirh}
$$
where $g_s=2$ is the supersymmetry generated $g$-factor, corresponding to
$\alpha^+=\alpha^-={e\over2M}$ whenever Eqn.~(\eqazoicp) holds%
\note{In the more
general case $g_s$ may be determined in terms of the eigenvalues of the
magnetic dipole matrix, and would take on two different values, with the
four $(j+1/2,j-1/2)$ pairs splitting into two plus two pairs.}.
Following the same analysis we find for the $j\times 1$ combination
$$
g_{j+1}=g_{j}+{{g_{s}-g_{j}}\over {j+1}}, \qquad
g_{j-1}=g_{j}-{{g_{s}-g_{j}}\over {j}}, \qquad
g_{j'}=g_{j}+{{g_{s}-g_{j}}\over {j(j+1)}}.
\numbereq\name{\eqamanati}
$$
Of the 5+1 states of spin $j$, five have a $g$-factor of $g_j$, while the
last has a $g$-factor of $g_{j'}$.  This demonstrates in particular that in
extended supersymmetry not all states of the same spin have to have the same
gyromagnetic ratio.

Next we turn our attention to the transition magnetic dipole moments.
This time, using the Wigner-Eckart theorem to write
$$
\langle j-\half, m+\half|{\mu_3}|j+\half, m+\half\rangle
={e\over 2M}h_{j}{\sqrt{j(j+1)-m(m+1)}},
\numbereq\name{\eqwrao}
$$
we find the transition elements
$$
h_{j}={{g_{j}-g_{s}}\over {2j+1}},\qquad
h_{j+{1\over 2}}={\sqrt{j\over {2j+1}}}{{g_{j}-g_{s}}\over {j+1}},
\qquad
h_{j-{1\over 2}}={\sqrt{{j+1}\over {2j+1}}}{{g_{j}-g_{s}}\over {j}},
\numbereq\name{\eqbantovi}
$$
where $h_{j\pm{1\over 2}}$ corresponds to
the matrix elements of the dipole moment operator
between the states with spins $j$ and $j\pm1$.

Short representations of $N=2$ have $g$-factors given by $g_{j\pm{1\over2}}$
in Eqn.~(\eqdsirh) and a transition moment given by $h_j$ in
Eqn.~(\eqbantovi).  Note that this agrees with the sum rule found in [\por]
as is expected due to the correspondence of $N=2$ short representations with
massive $N=1$ representations.


\newsection Conclusions.

In the above we have derived model independent sum rules for the
gyromagnetic and gyroelectric ratios of particles which
belong to a single $N=2$ supermultiplet.  As demonstrated in
Eqns.~(\eqdsirh), (\eqamanati) and (\eqbantovi), the gyromagnetic ratio of
any state in a generic long multiplet may be expressed in terms of the
quantities $g_j$ and $g_s$, where $g_s=2$ whenever the natural relation of
Eqn.~(\eqazoicp) holds.  Although we have examined Eqn.~(\eqazoicp)
carefully, we have as yet been unable to determine its validity in a
model-independent manner.  This leads us to believe that there may indeed be
models where ${\cal S}+i{\cal P}$ are free, thus allowing in addition a
$N=2$ supersymmetry contribution to the electric dipole moment, as indicated
in Tables 3 and 4.  This novel feature is somewhat surprising in that one
would usually anticipate the addition of more symmetry in going from $N=1$
to $N=2$ to lead to more restrictions and thus stronger sum rules on the
dipole moments.  However this is the opposite of what is actually found
above.  Furthermore, there is no contradiction with the sum rule determined
for the $N=1$ subalgebra of $N=2$.  In particular,  noting that
$$
\left|\up\right\rangle_{N=1} = {1\over\sqrt{2}}
\big[\sqrt{1+{\textstyle{|Z|\over2M}}}\left|\up\mneg\zip\right\rangle
-\sqrt{1-{\textstyle{|Z|\over2M}}}\left|\zip\mneg\up\right\rangle\big],
\numbereq
$$
we find the expected result
$\left\langle\up\right|\mu^3 \left|\up\right\rangle_{N=1}
= \mu_0+{e\over2M}$ independent of ${\cal S} + i{\cal P}$.

Just as the short representation of $N=2$ is connected with the massive
$N=1$ representation, the short representation of $N=4$ (preserving half of
the supersymmetries) is connected with the long representation of $N=2$.
Thus we expect the sum rules derived herein to also apply to the short $N=4$
case.  Since the latter were studied in [\jim], we may contrast the two
approaches.  While the present derivation is quite general, and focuses
on a conserved current commuting with supersymmetry, the latter took an
explicit $N=4$ (on-shell only) supergravity coupled Yang-Mills theory and
studied the dipole moments of BPS states through their asymptotic field
behaviours (although black holes were studied in [\jim], the sum rules
were derived in a general fashion, and depend only on having an appropriate
supersymmetric field configuration).  The resulting sum rule found in
[\jim] corresponds to the present $N=2$ long case, with $g_j=0$ and $g_s=2$.
In particular, this indicates that the $g$-factors for the $N=4$ short case
are completely fixed, hinting at the possibility that stronger sum rules do
arise in $N=4$ and $N=8$ theories (where the latter only has graviphotons)
that are not yet apparent in the $N=2$ case.

Finally, note that the technique of [\jim] allows a determination of
$g$-factor sum rules for graviphotons, which is not possible in the present
framework (since graviphotons do not commute with supersymmetry).  It would
be interesting, however, to see if model independent sum rules for
graviphoton couplings could be determined by working with a supergravity
multiplet instead of a real linear multiplet.


\newsection Acknowledgments.

We would like to thank V. P.~Nair and M.~Porrati for useful discussions.
This work was supported in part by the Department of Energy under Contract
Number DE-FG02-91ER40651-TASKB.

\immediate\closeout1
\bigbreak\bigskip

\line{\twelvebf References. \hfil}
\nobreak\medskip\vskip\parskip

\input refs

\vfil\end

\bye